\documentstyle[a4,12pt]{article}

\begin{document}
\begin{center}
{\bf FIRST SINGLE BUBBLE SONOLUMINESCENCE IN DUBNA.}
\end{center}
\begin{center}
V.B. Belyaev$^*$, M.B. Miller$^{**}$, A.V. Sermyagin$^{**}$\\
\end{center}
$^*$Osaka University, Mihogaoko 10--1, Ibaraki Osaka 567, Japan;
(on leave of Joint\\ Institute for
Nuclear Research, Dubna, Russia); {\it belyaev$@$miho.rcnp.osaka--u.ac.jp};\\
$^{**}$Institute in Physical--Technical
Problems, 141980 POB 39, Dubna, Moscow region,\\ Russia;
{\it sermyagin$@$vxjinr.jinr.ru}\\

At the Institute in Physical--Technical Problems experiments on
sonoluminescence was started by our group at the beginning of
this year. Our  study was  focused at properties
of the single bubble mode of sonoluminescence.  First
experiments have already taken us, as it seems, to
claiming--attention results.  In particular, we managed to
manipulate the process, varying boundary  conditions on
resonator outer surface. Some kind of interaction of a few
spaced bubbles was observed, enhancing light emission by all of them
 simultaneously.  Now we try to specify this
effect (``few bubble sonoluminescence'' -- FBSL)  and use it to
affect conditions within the bubble, and study it further.
Besides, conditions to induce SBSL in the spherical resonator with very low
 acoustical quality factor Q were discovered. This result,
somehow  unexpected, also attracted our attention. In such
conditions stable SBSL can  easy be  supported
during long-time runs. The system turns out to be practically
non-sensitive to changes in external environment.  The above
and some other results are presented in the report. Besides,
experiments in a stage of preparation are discussed, -- a study of
mechanism of  ``Argon rectification'' within the
bubble under SBSL, measurement of  statistical
moments of light flash intensity distribution, experiments on
Einstein--Podolsky--Rosen correlations, as well as experiments on nuclear
aspects  in sonoluminescence. The last direction of studies is
related, in  particular, to a possibility of ``molecular--nuclear
synthesis'' reactions in  water  under certain
conditions, much discussed recently [1].
\vskip 0.5cm
This work was supported in part by the Russian Foundation for Basic
Research, Grant No.\ 98--02--16884.


\begin{thebibliography}{10}
\bibitem{} V.B.Belyaev, A.K. Motovilov, W. Sandhas.
Can Water ``Burn''?  Physics Doklady, 1996, v.41, pp. 514--516.
\end{thebibliography}
\end{document}